\documentclass[prb,twocolumn,showpacs,superscriptaddress]{revtex4}

\usepackage{graphicx}
\usepackage{amsfonts}
\usepackage{amssymb}
\usepackage{framed, color}

\begin{document}

\title{Tailoring the frictional properties of granular media}

\author{Sonia Utermann}
\author{Philipp Aurin}
\author{Markus Benderoth}
    \affiliation{Max Planck Institute for Dynamics and Self-organization, Am Fa{\ss}berg 17, 37077 G\"{o}ttingen, Germany}
\author{Cornelius Fischer}
    \affiliation{Abt. Sedimentologie/ Umweltgeologie, GZG, Georg-August-Universit\"at G\"ottingen, Goldschmidtstra{\ss}e 3, 37077 G\"ottingen, Germany}
\author{Matthias Schr\"oter}
\email{matthias.schroeter@ds.mpg.de}
    \affiliation{Max Planck Institute for Dynamics and Self-organization, Am Fa{\ss}berg 17, 37077 G\"{o}ttingen, Germany}

\date{\today}


\begin{abstract}

A method of modifying the roughness of soda-lime glass spheres is presented, with the purpose of tuning inter-particle friction. The effect of chemical etching on the surface topography and the bulk frictional properties of grains is systematically investigated. The surface roughness of the grains is measured using white light interferometry and characterised by the lateral and vertical roughness length scales. The underwater angle of repose is measured to characterise the bulk frictional behaviour. We observe that the coefficient of friction depends on the vertical roughness length scale.


\end{abstract}


\maketitle

\section{Introduction}\label{introduction}

Experiments have confirmed that friction plays an important role in the physics of static granular materials. It was demonstrated that the random loose packing volume fraction decreases with increasing inter-particle friction coefficient \cite{Jerkins, Menon}. The role of friction on force chains was investigated in reference \cite{Blair}. If static granular media is described by a statistical mechanics approach \cite{Edwards}, a ``configurational temperature'' can be defined, which depends on friction \cite{Schroeter05}.

Friction plays a vital role in granular dynamics, too; for instance in the case of granular material undergoing shear \cite{Tamas, Mair, Anthony}. The effect of friction in granular materials under shear in a linear, split-bottomed shear cell filled with layers of high-friction and low-friction grains has been investigated \cite{Tamas}. Here, it was shown that the position of the shear band depends on the layering of the two types of grains. It has been shown experimentally that friction can also play an important role in granular segregation \cite{Plantard, Ulrich}, for example in the sharp transition from the reverse brazil-nut effect to the brazil-nut effect \cite{Ulrich}, but other work shows that this is not always the case \cite{Pohlman}. The dynamical properties of avalanches were shown to depend on inter-particle friction, and two friction-dependent scenarios for granular avalanches on an incline were identified \cite{borszonyi_1, borszonyi__2, borszonyi___3}.

\begin{figure}
\includegraphics[width=8cm]{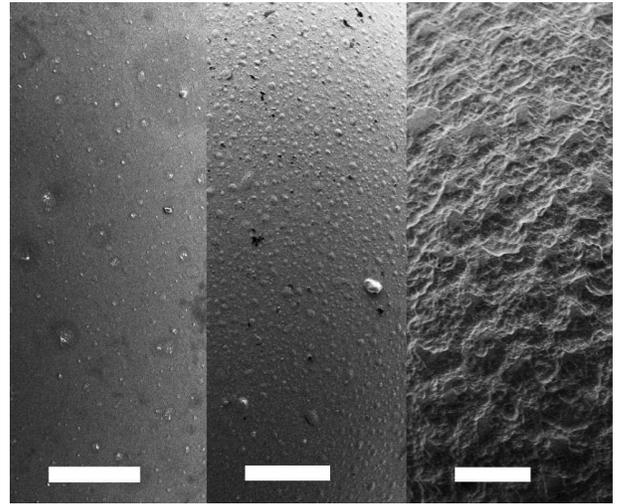}
\caption[dum]{\label{SEM_photomontage} \small\ SEM images of the surfaces of MoSci glass spheres of diameter 146 $\mu$m $\pm$ 19 $\mu$m (small Mo-Sci). Left: smoothed by NaOH. Centre: untreated. Right: etched for four minutes. The white bars are 10 $\mu$m long.}
\end{figure}

However, none of the above work was made with particles of the same material and shape but more than two coefficients of friction. In this study, it is shown that it is possible to tune the topographic properties of soda-lime glass grains, making inter-particle friction a control parameter for some types of beads and thus allowing the systematic quantitative study of the principles underlying granular phenomena such as those mentioned above. Figure \ref{SEM_photomontage} shows the surfaces of grains when they are untreated and after etching with the two protocols we describe below.\par

The article is structured as follows: in section \ref{etching protocol}, the chemical etching protocols are described. The analysis of the grain roughness follows in section \ref{roughness methodology}. In section \ref{friction}, we present the bulk frictional properties of the grains measured by the underwater angle of repose and in appendix \ref{Coke} a bulk measure of roughness is presented.\par

\section{Etching glass grains}\label{etching protocol}

\begin{figure}[here]
\begin{center}
\includegraphics[width=6cm]{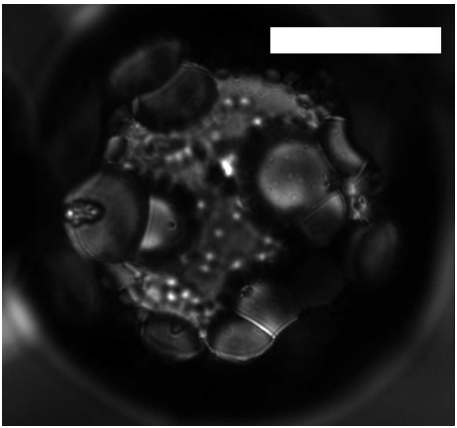}
\caption[dum]{\label{HF_grain} \small\ Optical micrograph of a glass sphere etched in HF. The white bar is 50 $\mu$m long. Instead of homogeneous surface roughening, large craters are produced.}
\end{center}
\end{figure}

Usually, hydrofluoric acid (HF) is used to etch glass \cite{Blair, Spierings}. However, HF alone is unpleasant to work with and is harmful to the environment \cite{MSDS_HF}. Furthermore, our experiments with dilute HF showed highly inhomogeneous etching unconducive to our purpose, as demonstrated in figure \ref{HF_grain}. Ammonium bifluoride showed better etching properties (figure \ref{SEM_photomontage}) as well as better handling: an aqueous solution of ammonium bifluoride will contain only small amounts of free HF, according to the set of equilibrium equations \ref{HF}.

\begin{eqnarray}\label{HF}
    \mathrm{NH}_4\mathrm{FHF} \rightleftharpoons \mathrm{NH}_4^{+} + \mathrm{FHF}^{-} \rightleftharpoons \mathrm{NH}_4^+ + \mathrm{HF} +\mathrm{F}^-\\
    3\mathrm{HF} + 2\mathrm{H}_2\mathrm{O} \rightleftharpoons \mathrm{FHF}^- +\mathrm{F}^- + 2\mathrm{H}_3\mathrm{O}^+\nonumber
\end{eqnarray}

A protocol was developed using ammonium bifluoride which optimises the flow properties of the etch solution whilst releasing little hydrofluoric acid and etching as homogeneously as possible \cite{etch patent}.\par

\subsection{Roughening protocol}

This is a hydrofluoric acid etching procedure, so the utmost care must be observed during all steps in order to avoid chemical burns. Since the procedure is one to etch glass, plastic lab ware should used.\par

\begin{enumerate}
\item 14 parts by mass of liquid glycerine, $\rm C_3H_8O_3$, and six by mass of distilled water are heated in a water bath to $90^{\circ}$C.
\item Six parts by mass of solid ammonium hydrogen-difluoride, $\rm NH_4HF_2$, and one part by mass of solid granulate iron(III)chloride-hexahydrate, $\rm FeCl_3\cdot 6H_2O$ (a catalyst), are added.
\item An over-saturated solution is produced by mixing vigorously at $90^{\circ}$C.
\item The solution is cooled under agitation until it is cool enough to be placed in the centrifuge.
\item The over-saturated solution is centrifuged until the clear solution is separated from the undissolved solute (about 15 minutes).
\item The solution is added to the glass grains and shaken vigorously. After the desired etching time has passed, the solution is poured out and the grains thoroughly washed with distilled water.
\end{enumerate}

A separate etching procedure to make the grains smoother was also employed. This method was published by Schellenberger and Logan \cite{Schellenberger}, To smoothen the grains, they are immersed in 12.5 M sodium hydroxide, $\rm NaOH$, for 30 minutes, under occasional agitation.\par



We use soda-lime glass grains from two different manufacturers and in two different sizes. These are presented in the table below.\par

\begin{table}[h]
{\label{grainstable}}
\begin{tabular}{|c|c|c|c|} \hline
    Name & Diameter ($\mu$m) & STD ($\mu$m)& Symbol\\ \hline
    Small Mo-Sci\cite{mo-sci} &  146 & 19 & $\color{blue}{\blacktriangle}$ \color{black}\\
    Cataphote\cite{cataphote} & 221 & 40 & $\square$\\
    Large Mo-Sci &  240 & 20 & $\color{red}{\bullet}$\color{black}\\ \hline
\end{tabular}
\caption[dum]{\label{table} \small\ (Colour online) The types of grains used in this study. The particle diameter and standard deviation (STD) was measured using a Camsizer, as outlined below.}
\end{table}

We perform seven treatments on each type of glass grain: smoothed for 30 minutes in NaOH, as supplied by the manufacturer (``unetched'') and roughened for 30 seconds and one, two, three and four minutes.\par


The effect of etching on particle size was measured using a \emph{Retsch Technology} Camsizer. The particle size distribution of 200,000 grains of each batch was measured. The reduction in grain diameter after four minutes' etching with the roughening protocol is very small: between unresolvable within experimental error for the large Mo-Sci to a reduction of 3.5 $\mu$m $\pm$ 1.5$\mu$m (a reduction of 2.4\%) for the small Mo-Sci grains. The smoothing protocol does not bring about a resolvable reduction in diameter. The width of the size distributions also remains constant within experimental error.\par


\section{Grain roughness}\label{roughness methodology}

\begin{figure}[here]
\begin{center}
\includegraphics[width=6.5cm]{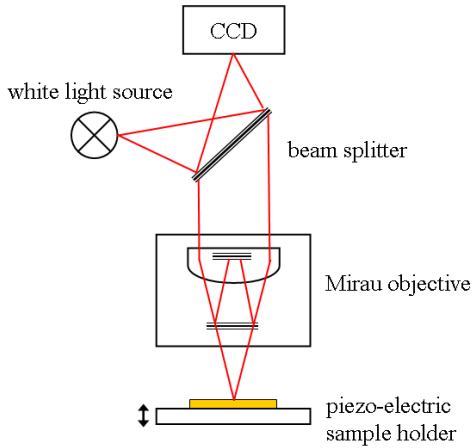}
\caption[dum]{\label{VSI_principle_of_function} \small\ (Colour online) Principle of operation of a vertical scanning interferometric microscope.}
\end{center}
\end{figure}

\begin{figure}[here]
\begin{center}
\includegraphics[width=9cm]{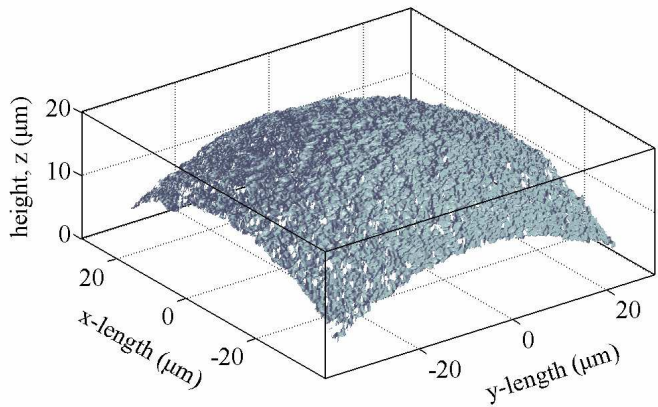}
\caption[dum]{\label{WLI_example} \small\ Height map of the cap of a large Mo-Sci grain etched for four minutes.}
\end{center}
\end{figure}

The grain surface topography was measured using white-light vertical scanning interferometry (WLI), (ZeMapper, \emph{Zemetrics} Inc., Tucson, Arizona) \cite{Dharba}. WLI is an optical surface scanning method that is able to provide nm height resolution over a large field of view. This enables the characterisation and quantification of surface topography variations of, for example, natural grains, reacted crystal surfaces, and mineral aggregates \cite{Fischer}. The principle of operation is shown in figure \ref{VSI_principle_of_function}. After passing through an external beam splitter, the white light passes through a temperature-stabilised Mirau objective (magnification: 160 x). The objective is equipped with a second, internal, beam splitter. One beam is reflected by a stationary reference mirror, the other by the sample surface. The two light beams differ in optical path length as a function of the distance to the sample surface. The resulting interference pattern is captured by a high-resolution charge-coupled device (CCD). The focal plane of the interference pattern is scanned through the sample. From the collected interference pattern data, a height map is calculated. The vertical resolution of the data set is $<$1 nm; the maximum field of view applied in this study is 80 $\mu$m x 80 $\mu$m (figures \ref{VSI_principle_of_function} and \ref{WLI_example}).\par

For the WLI measurements, glass grains are bonded to glass microscope slides with ultra-violet-curable adhesive (\emph{Norland} Optical Adhesive 61) and sputter-coated with 40 nm gold. \par

\subsection{Roughness analysis}\label{The meaning of roughness}

\begin{figure}
\begin{center}
\includegraphics[width=5cm]{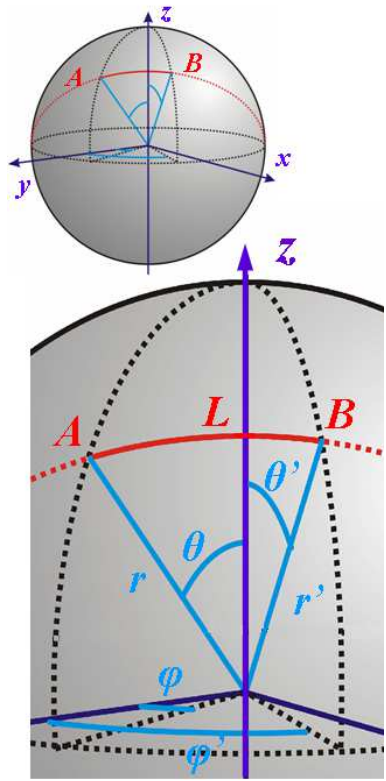}
\caption[dum]{\label{explaining roughness} \small\ (Colour online) The method employed to measure roughness takes account of the spherical topology of the grain. The height difference correlation, $\rho (L)$, between pairs of points $r(\theta,\phi)$ and $r'(\theta',\phi')$, separated by their great-circle distance, $L$, is calculated. $R$ is the radius of the grain.}
\end{center}
\end{figure}

Characterising roughness alone is a complex endeavour, and there is a variety of tools available to do so \cite{roughness}. We use the height difference correlation function \cite{Meakin}. However, when measuring the roughness of spheres, one is presented with additional, unique problems. To obtain reliable statistics, a large field of view is necessary. However, the curvature of the grain means that the measurement method must have a large enough dynamic range to be able to measure nano-scale features over a height range of several $\mu$m. WLI proved to be better than AFM for this purpose.\par

Once a height map, $z(x,y)$, has been obtained, this may not be treated as a plane, since it is a projection of a sphere surface onto a plane, which can never be simultaneously conformal and distance conserving \cite{Bronstein}. It is also necessary to make an angle-of-sight correction to account for having measured height in the $z$-co-ordinate instead of in the radial direction. This correction is explained in appendix \ref{angle-of-sight correction}. Additionally, a batch of soda-lime glass grains will not consist only of perfectly spherical grains.\par

For each batch of grains, ten to fifteen grains are analysed. From the surface relief measurement, the typical surface feature size characterised by a vertical, $\xi_{vert}$, and a lateral length scale, $\xi_{lat}$, is measured.\par

For a one-dimensional interface, $\xi_{vert}$, $\xi_{lat}$ and the Hurst exponent, $H$, are obtained from the (second-order) height difference correlation function \cite{Meakin}

\begin{equation}
C(L) = \langle (h(x+L)-h(x))^2\rangle ^{1/2},
\end{equation}

where $h$ is the height of the interface at position $x$ or $x + L$. For the surface of a rough sphere described in spherical polar co-ordinates, the height difference correlation becomes

\begin{equation}
\rho (L)=\langle (r'(\theta',\phi')-r(\theta,\phi))^2 \rangle^{1/2},
\end{equation}

where the points $r(\theta,\phi)$ and $r'(\theta',\phi')$ are separated by the correlation distance $L$. On a sphere, $L$ is the great-circle distance between the two points calculated using the spherical cosine law:

\begin{equation}
L = R \arccos (\cos\theta\cos\theta'+\sin\theta\sin\theta'\cos(\phi'-\phi)),
\end{equation}

where $R$ is the sphere radius. The angle brackets indicate an ensemble average for one $L$.

\begin{figure}[here]
\begin{center}
\includegraphics[width=8cm]{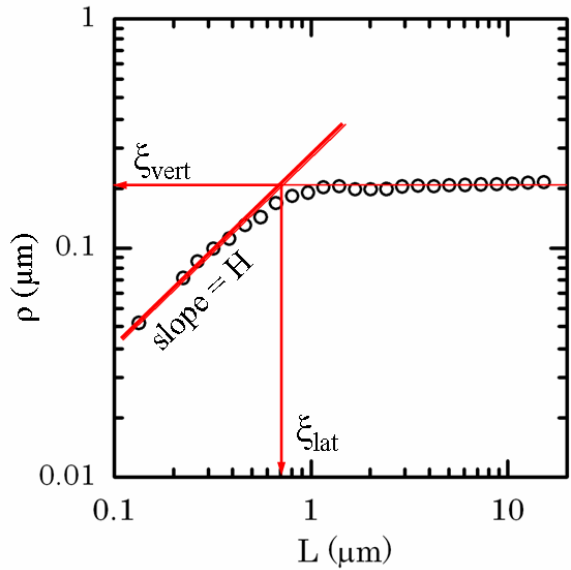}
\caption[dum]{\label{R_of_L_explanation} \small\ (Colour online) The roughness, $\rho (L)$, on a double logarithmic scale, yields the vertical and lateral saturation length scales, $\xi_{vert}$ and $\xi_{lat}$, here for a Cataphote grain etched for 30 seconds. The slope of the fit through the first five data points is the Hurst exponent, $H$.}
\end{center}
\end{figure}

The height difference correlation, $\rho (L)$, on a double logarithmic scale, yields the vertical and lateral saturation length scales, $\xi_{vert}$ and $\xi_{lat}$, for a given etch time. This is the position at which a cross-over between a power-law type behaviour and a saturation plateau occurs, and is illustrated in figure \ref{R_of_L_explanation}.\par

The roughness length scales are obtained from the WLI micrographs of the spheres by the following steps: first, the WLI measurement is performed over a field-of-view of 80 $\mu$m $\times$ 80 $\mu$m (60 $\mu$m $\times$ 60 $\mu$m for the small Mo-Sci). The radius of the grain, $R$, is measured by optical micrography. The data is cropped to 60 $\mu$m $\times$ 60 $\mu$m (30 $\mu$m $\times$ 30 $\mu$m) to remove the bad pixels at the edges, then processed to remove spurious data adjacent to known bad pixel sites. Bad pixels make up less than 1\% of the cropped data and they are mostly at the edges where the effect of grain curvature is highest. Next, the measured $R$ and an estimate of the $x$ and $y$ positions of the ``north pole'' are taken as start values for a hemispherical fit to the data. Using the position of the centre of the grain, $x_0, y_0$ and $z_0$, obtained from the hemispherical fit, the data is converted to spherical polar co-ordinates $r, \theta$ and $\phi$.\par

Following this, $\rho (L)$ is calculated, taking account of the angle-of-sight correction (appendix \ref{angle-of-sight correction}). Three million randomly-chosen pairs of points are taken to calculate $\rho (L)$ for each grain. The vertical length scale, $\xi_{vert}$, is obtained by laying spline through $\rho (L)$; the length $L_{kink}$ at which the slope of this spline becomes zero is detected. $\rho (L)$ is averaged for $L > L_{kink}$, giving $\xi_{vert}$. A linear fit through the first five data points gives the Hurst exponent, $H$. The cross-over of the linear fit and the plateau ($\xi_{vert}$) gives $\xi_{lat}$.

Strong asphericity in the grain will lead to distortion of the roughness analysis, since our analysis assumes a spherical topology. To establish the quality of the data, radial averaging is performed (see appendix \ref{radial_averaging}) and data from aspherical grains is discarded. Of the fifteen grains measured per batch, on average two were rejected for asphericity.\par

\subsection{Roughness results}\label{roughness results}

\begin{figure}[here]
\begin{center}
\includegraphics[width=11cm]{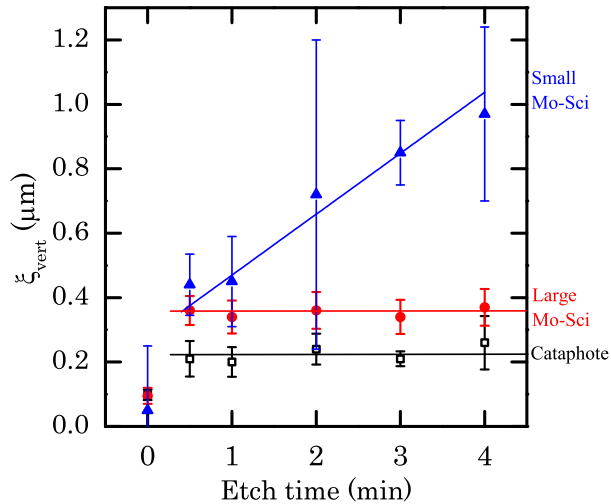}
\caption[dum]{\label{xivert} \small\ (Colour online) The mean vertical roughness length scale, $\xi_{vert}$, as a function of $\mathrm{NH}_4\mathrm{FHF}$ etch time. In this and all subsequent plots, the error bars are the standard deviation; for each grain type ten to fifteen grains were measured. The linear fit through small Mo-Sci points and the means of $\xi_{vert}$ are in the range of 30 s to 4 min.}
\end{center}
\end{figure}

\begin{figure}[here]
\begin{center}
\includegraphics[width=9cm]{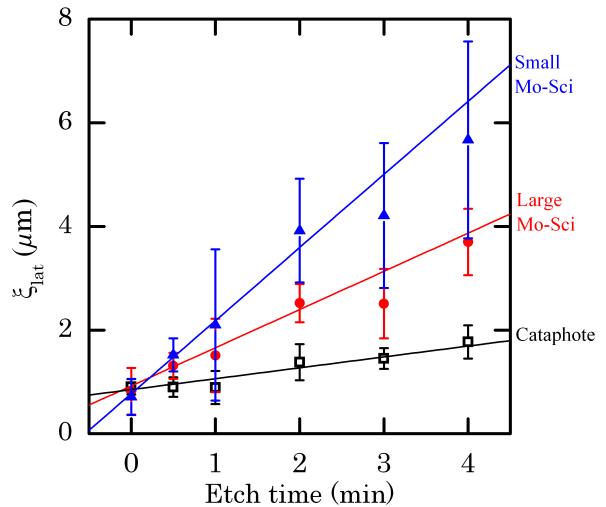}
\caption[dum]{\label{xilat} \small\ (Colour online) The mean lateral roughness length scale, $\xi_{lat}$, as a function of $\mathrm{NH}_4\mathrm{FHF}$ etch time.}
\end{center}
\end{figure}

The effect of etching on the topography of glass grains is shown in figures \ref{xivert} and \ref{xilat}, where the mean vertical and lateral roughness length scales, $\xi_{vert}$ and $\xi_{lat}$ respectively, are plotted as a function of $\mathrm{NH}_4\mathrm{FHF}$ etch time.\par

From figures \ref{xivert} and \ref{xilat}, one sees that there are two types of roughening etching behaviour: a monotonic increase in feature width and depth with etch time for the small Mo-Sci grains, where $\xi_{vert}$ increases at an approximate rate of 0.17 $\mu$m per minute and $\xi_{lat}$ at a rate of 1.4 $\mu$m per minute; and a broadening behaviour for the two larger types of grains, whereby surface features become \emph{wider} but not \emph{deeper} as a function of etch time. It is not clear from our data if the change in etch mechanism is connected to the different radii or changes in the glass composition. The smoothed grains show no appreciable difference to the unetched grains, and are omitted from the plots.\par

\begin{figure}[here]
\begin{center}
\includegraphics[width=9cm]{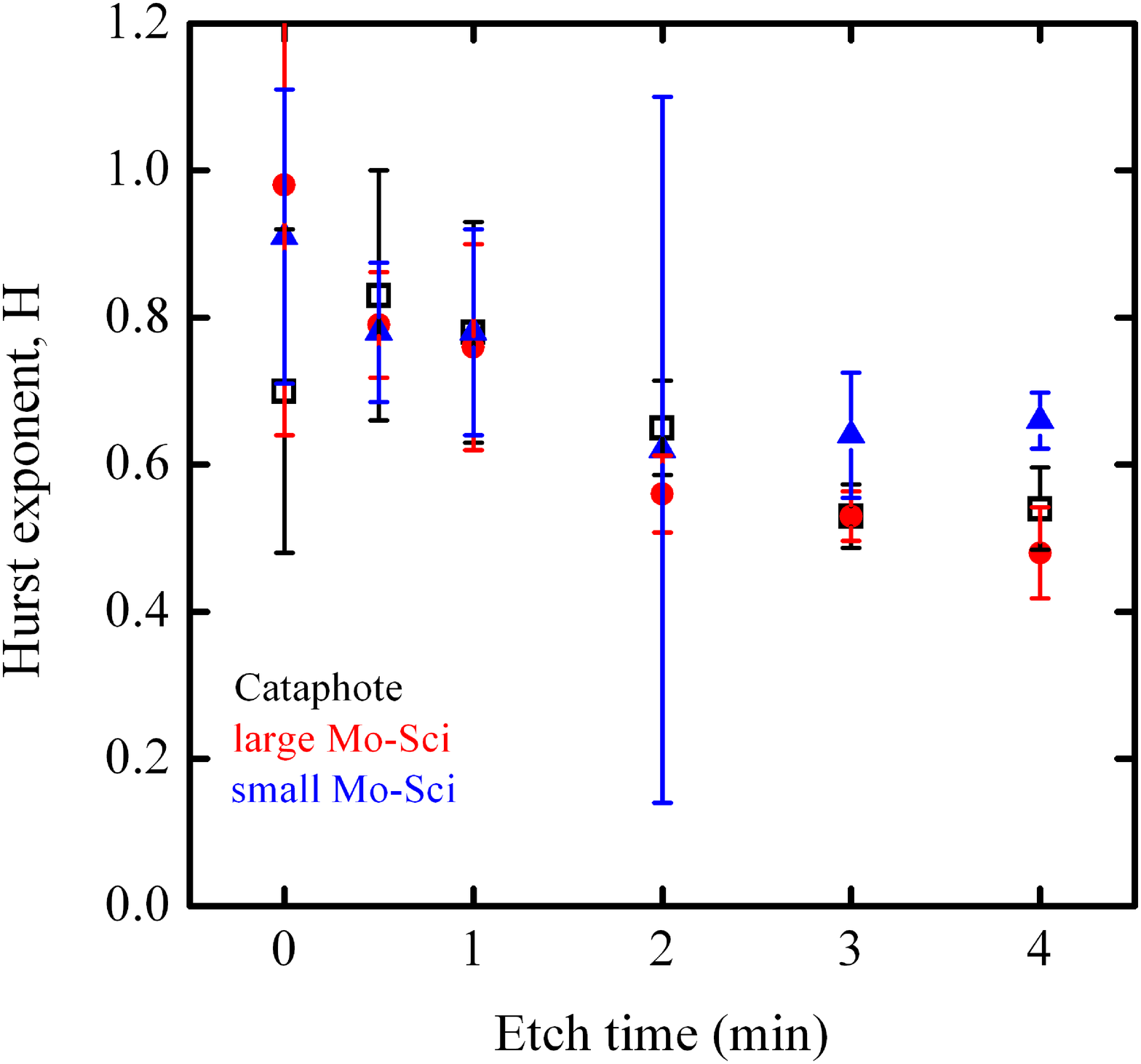}
\caption[dum]{\label{Hurst} \small\ (Colour online) The mean Hurst exponent, $H$, as a function of $\mathrm{NH}_4\mathrm{FHF}$ etch time.}
\end{center}
\end{figure}

The Hurst exponent may also influence grain interactions \cite{Halsey}. However, as can be seen from figure \ref{R_of_L_explanation}, the power-law regime of $\rho (L)$ exists over less than an order of magnitude, so caution must be exercised in interpreting the behaviour of the Hurst exponent, $H$, as a function of etch time (figure \ref{Hurst}). We see $0 < H \leq 1$ for all 21 grain types. $H$ shows a slight downwards trend with increasing etch time which is the same for all three grain sets, suggesting that the Hurst exponent can also be tuned by the etching protocols presented here.

It is desirable to be able to measure surface roughness in bulk samples. A possible method for doing this,  using a carbonated soft drink, is demonstrated in appendix \ref{Coke}.


\section{Bulk frictional properties}\label{friction}

In this section, the angle of repose, $\alpha$, is measured to characterise the bulk frictional behaviour; to eliminate electrostatic and capillary attraction between grains, the measurements are made under water.

\subsection{Angle of repose measurements}\label{angle of repose}

\begin{figure}[here]
\begin{center}
\includegraphics[width=8cm]{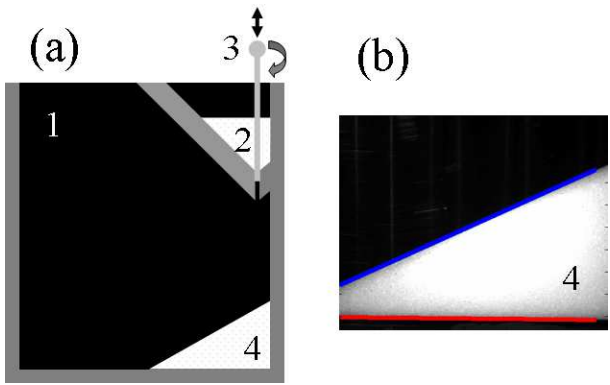}
\caption[dum]{\label{AOR_set-up} \small\ (Colour online) Angle of repose measurement method. (a) A sketch of the set-up: a cell [1] 25 cm wide, 30 cm high and 40 grain diameters deep is filled with water. The reservoir [2] is filled with the granular sample. The funnel rod [3] is raised by a screw thread to open the hole enough to let grains flow to the bottom part of the cell [4]. After an avalanche has occurred, the rod is lowered enough to induce granular arching, thus stopping the flow of grains. (b) CCD image of the pile. Linear fits to the bottom and top edges of the pile give the angle of repose (see text).}
\end{center}
\end{figure}

The angle of repose of each sample is measured using the set-up described in figure \ref{AOR_set-up}. A cell 25 cm wide, 30 cm high and 40 grain diameters deep is filled with water \cite{Footnote_midi}. The water temperature is kept at $21 \pm 0.5^{\circ}$C for all experiments to keep the viscosity constant. An upper reservoir is filled with the granular sample. Then grains are allowed to flow to the bottom part of the cell at a rate of approximately 140 grains/s. After an avalanche has occurred, the flow of grains to the pile is stopped. The pile is photographed with a black-and-white CCD with 1 Megapixel and 8-bit depth. Each photograph is binarised using a threshold that is chosen automatically to fall between the peaks of its bimodal grey-level histogram. The top and bottom edges of the pile are detected and the location of the linear part of the pile is detected by laying a spline through the line describing the top of the pile. Within this range, linear fits are made to the top and bottom of the pile; the angle between them is the angle of repose, $\alpha$. The measurement was repeated ten to twenty times per grain type. The resolution in $\alpha$ is $\pm 0.15^{\circ}$\par

\subsection{Results}

\begin{figure}[here]
\begin{center}
\includegraphics[width=9cm]{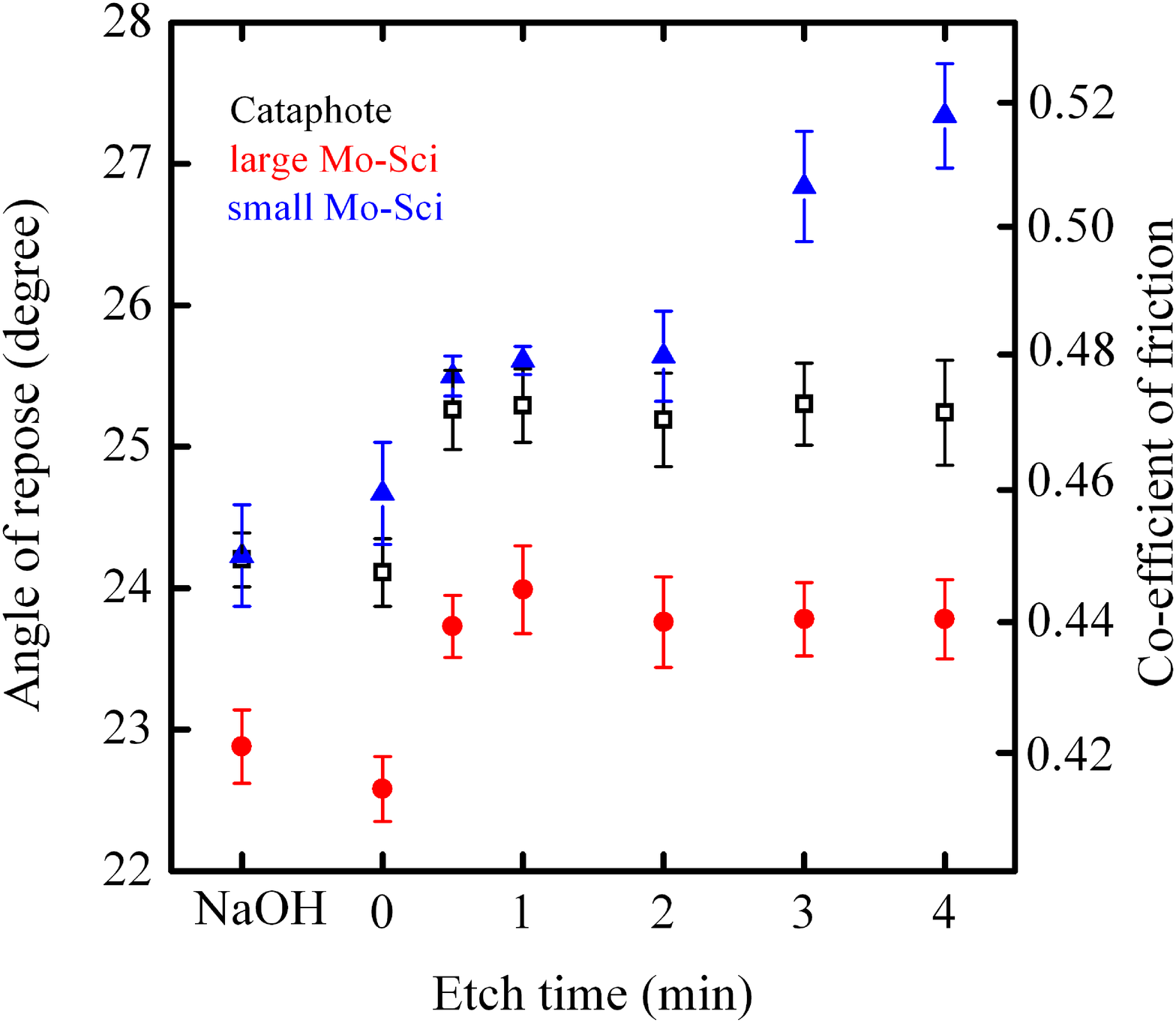}
\caption[dum]{\label{AOR_vs_etch_time} \small\ (Colour online) Underwater angle of repose as a function of roughening etch time. The smoothing protocol, NaOH, is plotted on the negative $x$-axis. The error bars are the standard deviation. The coefficient of friction is computed by $\mu = \tan \alpha$, where $\alpha$ is the angle of repose \cite{Coulomb}.}
\end{center}
\end{figure}

Figure \ref{AOR_vs_etch_time} shows $\alpha$ as a function of roughening etch time. For the small Mo-Sci grains, the angle of repose increases monotonically from about $24^{\circ}$ to over $27^{\circ}$ as a function of etch time. As in the roughness measurements, the smoothed grains do not show an appreciable difference to the unetched grains.\par

The Cataphote and large Mo-Sci grains again show a different behaviour to the small grains. Again, their behaviour is a step function: unetched and smoothed grains have an angle of repose about one degree less than grains etched in $\mathrm{NH}_4\mathrm{FHF}$ for 30 s or longer. The appearance of these trends is similar to $\xi_{vert}$ as a function of etch time shown in figure \ref{xivert}. However, the Cataphote grains, whose $\xi_{vert}$ is always lower than the large Mo-Sci grains, have a consistently \emph{larger} angle of repose.

One reason for this discrepancy could be the different fractions of non-spherical grains present in the Cataphote and Mo-Sci samples. The average length-to-width ratio, $l/b_{ave}$, of the unetched Cataphote and large Mo-Sci grains was measured using a \emph{Retsch Technology} Camsizer (sample size of 7,000 grains). With $l/b_{ave} = 1.17$, the Cataphote grains show a slightly greater average asphericity than the large Mo-Sci grains, $l/b_{ave} = 1.12$. It has been demonstrated that higher asphericity leads to a larger angle of repose \cite{borszonyi___3}. Another reason for the discrepancy could be that the two larger types of grain also produce different packing densities, $\phi$, whish would be a natural consequence of their aspherity \cite{Donev}. Though we did not measure $\phi$ in this set-up, we show that $\phi$ has a strong influence on the angle of repose (appendix \ref{phi}). Again, for the larger grains, the smoothing protocol has no apparent influence on $\alpha$. \par

\begin{figure}[here]
\begin{center}
\includegraphics[width=9.5cm]{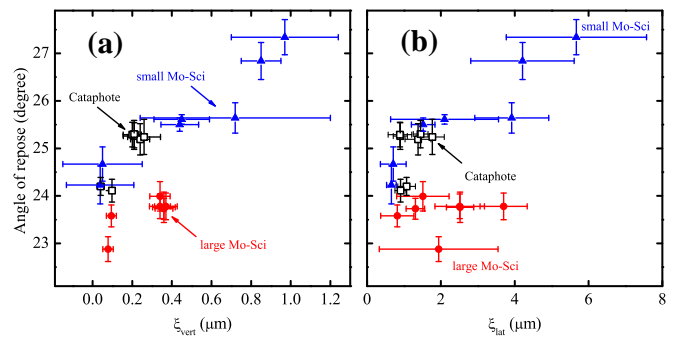}
\caption[dum]{\label{AOR_lengths} \small\ (Colour online) Angle of repose, $\alpha$, as a function of correlation length scales (a) $\xi_{vert}$ and (b) $\xi_{lat}$.}
\end{center}
\end{figure}

A valuable result of this work is to show how the frictional properties of grains depend on their (tailored) roughness. This is illustrated in figure \ref{AOR_lengths}: the angle of repose has a dependency on $\xi_{vert}$, albeit a weak/noisy one. The angle of repose showed no appreciable correlation with $\xi_{lat}$, again suggesting that it is the \emph{depth} of surface features, not their width, that controls the friction between grains. However, there is not a simple, linear dependency.

\section{Conclusion}\label{conclusion}

In this study, a method of modifying the topographical properties of soda-lime glass grains was presented, with the purpose of tuning inter-particle friction. We presented a roughening and a smoothing chemical etching procedure. The surface roughness of the grains was measured using white light interferometry and characterised using the height difference correlation function. Two types of roughening etching behaviour were observed: a monotonic increase in feature width and depth with etch time for the small grains; and a broadening behaviour for the two larger types of grains, whereby surface features became wider but not deeper as a function of etch time. We concluded that, when changing the roughness of glass grains, the type of grain matters. With the three types of grains we used, we are unable to distinguish whether this is due to size or material composition. The smoothing protocol made little difference to the surface roughness. The underwater angle of repose, $\alpha$, was measured to characterise the bulk frictional behaviour. For the small grains, $\alpha$ increased monotonically as a function of etch time. The two larger types of grains showed a different behaviour: unetched and smoothed grains had an angle of repose about one degree less than etched grains. In the appendix, we presented a suggestion for measuring grain roughness in bulk samples.\par

The ammonium bifluoride etching technique presented here allows the experimentalist to modify the roughness of glass spheres. Grains with tunable roughness are the ideal work-horse for experimental studies of wet (or ``humid'') granular systems \cite{Halsey} or of hydrodynamic interactions between grains in viscous fluids \cite{Beetstra,Jenkins}. It is shown that changing the surface roughness of grains directly changes the way they behave in the bulk. Besides the systems already shown experimentally to depend on friction, which were reviewed in the introduction, this has other important consequences for granular community in that it paves the way for comparison with simulations, where friction between grains is a typical control parameter. Additionally, the ability to tailor glass grains will allow systematic study of the role of friction in the static and dynamic behaviour of granular material, for instance granular segregation and pattern formation, avalanching, and (shear)flow, packing and force chains.


\begin{appendix}

\section{Angle-of-sight corrections}\label{angle-of-sight correction}

\begin{figure}[here]
\begin{center}
\includegraphics[width=6cm]{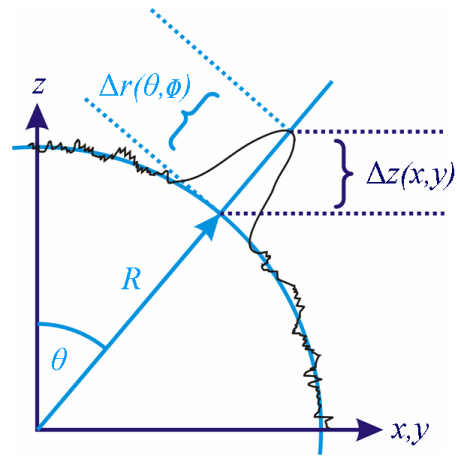}
\caption[dum]{\label{angle-of-sight_sketch} \small\ (Colour online) When analysing the roughness of the glass spheres, we need to make an angle-of-sight correction to account for our having measured height in the $z$-co-ordinate instead of in the radial direction. $R$ is the sphere radius and $\Delta z(x,y)=h(\theta,\phi)$ is the height of the grain profile measured vertically from above. The correct radial height of the feature is given by $\Delta r = \frac{\Delta z}{\cos\theta}$.}
\end{center}
\end{figure}

When analysing the roughness of the glass spheres, we make an angle-of-sight correction to account for our having measured height in the $z$-co-ordinate instead of in the radial direction. This is shown in figure \ref{angle-of-sight_sketch}. We correct the vertical height, $\Delta z(x,y)$, to a radial feature size, $\Delta r(\theta,\phi)$, by

\begin{equation}
\Delta r = \frac{\Delta z}{\cos\theta}
\end{equation}

The relevance of this correction can be estimated for the worst-case scenario: a feature at the outermost corner of the region of interest on a small Mo-Sci grain. In this case, $\theta=37^{\circ}$; thus omitting this correction results in an error of 26\%.

\section{Radial averaging to test for aspherical grains}\label{radial_averaging}

Strong asphericity in the grain will lead to distortion of the roughness analysis, since our analysis assumes a spherical topology. In our case, this is particularly true of the two smoothest sets of grains. To check if the grain is spherical enough that the roughness data is not falsified, radial averaging of the relief measurement of the grain is performed:

\begin{equation}
\langle \Delta r(\theta)\rangle|_{all \phi}.
\end{equation}

\begin{figure}[here]
\begin{center}
\includegraphics[width=8cm]{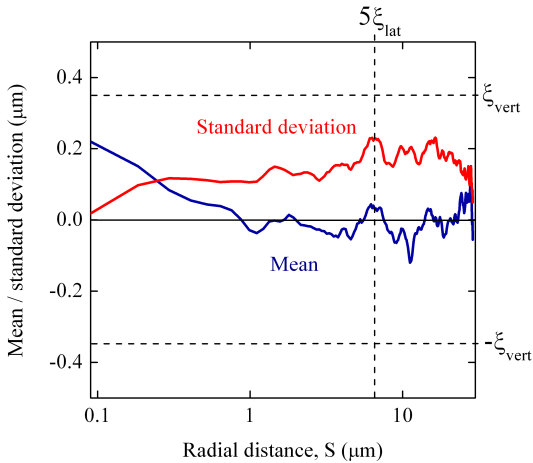}
\caption[dum]{\label{radial_good} \small\ (Colour online) Radial average and standard deviation of a large Mo-Sci grain etched for one minute. This grain is accepted: both lines remain within the range $[-\xi_{vert}...\xi_{vert}]$ for $L<5\xi_{lat}$.}
\end{center}
\end{figure}

\begin{figure}[here]
\begin{center}
\includegraphics[width=8cm]{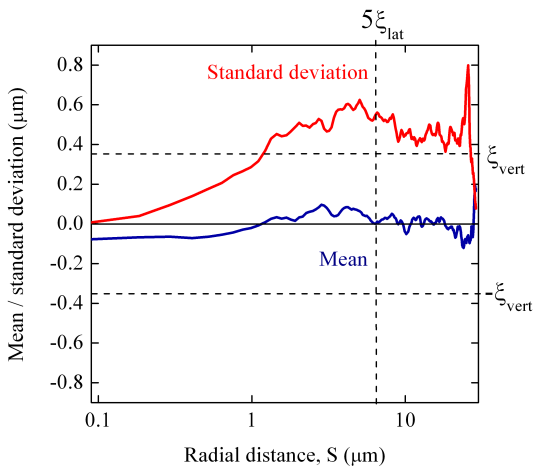}
\caption[dum]{\label{radial_bad} \small\ (Colour online) Radial average and standard deviation of a different large Mo-Sci grain etched for one minute. This grain is rejected: the standard deviation exceeds the range $[-\xi_{vert}...\xi_{vert}]$ for $L<5\xi_{lat}$.}
\end{center}
\end{figure}

For a perfect sphere, both the radial mean and the standard deviation are zero for all $\theta$. For rough grains, small deviations from zero are expected, for aspherical grains large deviations. Since our analysis depends on our ability to locate the kink in $\rho(L)$, it is vital that the mean and the standard deviation are smaller than the length scales we wish to measure, $\xi_{vert}$ and $\xi_{lat}$. We set the criterion that, if the radial mean OR standard deviation is greater than $\xi_{vert}$ for any angle corresponding to a radial distance $S<5\xi_{lat}$, we reject the grain from our analysis, on the grounds that its asphericity has an influence on its $\rho (L)$ at the relevant length scales. This is illustrated in figures \ref{radial_good} and \ref{radial_bad}.\par

Of the fifteen grains measured per batch, on average two were rejected for asphericity.


\section{Using a carbonated soft drink to measure the roughness of grains}\label{Coke}

The popular experiment in which a geyser is produced by dropping Mentos into a bottle of Diet Coke \cite{Mythbusters} was studied systematically by T. S. Coffey \cite{Coffey}. She concluded that the surface roughness of the sweets is one of the main causes of the reaction; namely, surface features provide nucleation sites for bubbles of carbon dioxide. We use this principle in a prototype experiment for measuring surface roughness in bulk granular samples using a carbonated soft drink.\par

The experimental method is as follows: 4 ml $\pm$ 0.05 ml of a freshly-opened carbonated soft drink (Coca-Cola) is poured into a measuring cylinder. Pouring is smooth to minimise premature nucleation of bubbles. 500 mg $\pm$ 0.5 mg of granular sample (small Mo-Sci) is added through a funnel. The maximum volume of cola, beads and foam is measured and the volume of the grains and the initial volume of cola subtracted.\par

\begin{figure}[here]
\begin{center}
\includegraphics[width=9cm]{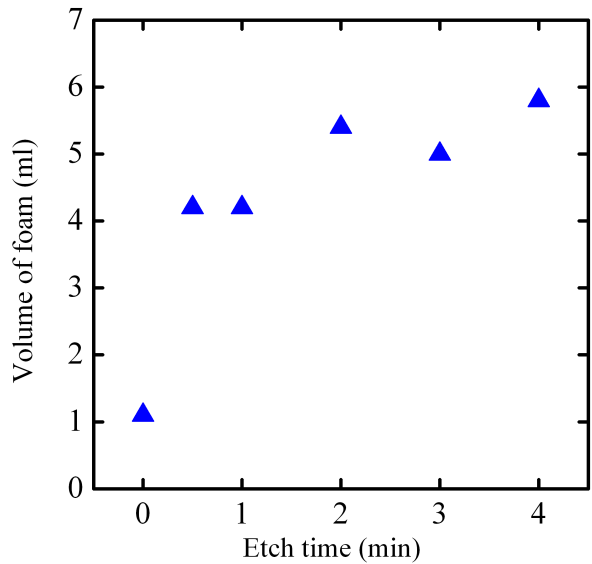}
\caption[dum]{\label{coke_foam} \small\ Maximum volume of cola foam as a function of etch time (small Mo-Sci).}
\end{center}
\end{figure}

Figure \ref{coke_foam} shows that there is a trend: the cola produces more foam when it comes into contact with the more strongly etched particles. Of course, not just surface roughness but also other effects, such as surface activity, buoyancy and occupation density of nucleation sites play a vital role in the reaction \cite{Coffey}. It is interesting to speculate on how this method could be calibrated.

\section{Angle of maximum stability as a function of packing fraction}\label{phi}

The angle of repose (pile slope after an avalanche, $\alpha$) is dependent not only on the friction between grains but on their packing density. The same is true of the angle of maximum stability (pile slope just before an avalanche, $\alpha_{max}$), which proved to be the more accessible quantity in investigating this dependence.\par

The density of the grain material is determined to within 0.45 $\mu$g/cm$^3$ with a \textit{Micromeritics} gas pyknometer. The grains are added to a cell of square cross-section 4 cm $\times$ 4 cm which is filled with water. A packing of grains is (re)produced on a Ling Dynamic Systems V409 shaker. The packing fraction is determined to within an error of $\pm 9\times 10^{-3}$ by measuring the height of the packing in the container. The container is transferred to a board which is hinged at one end and raised slowly at the other, under computer control, by a stepper motor. A black-and-white CCD with 1 Megapixel and 8-bit depth is used to record the surface of the granular sample while the angle of the container is increased. The camera and an LED light source are mounted such that they do not move relative to the container, thus avoiding spurious results from shadows; the camera is triggered as the stepper motor starts. Image processing is used to determine $\alpha_{max}$ with an accuracy of $\pm 0.5^{\circ}$ as follows: First, subsequent images of the surface of the sample are subtracted. The subtracted images are binarised; the threshold for binarisation is taken as the mean grey value for a non-moving bed plus four standard deviations. Then, connected moving pixels are detected. A continuous moving area of more than one grain (more than 50 pixels with our optics and set-up) is our definition of an avalanche. The angle at which this occurs is the angle of maximum stability, $\alpha_{max}$.\par

\begin{figure}[here]
\begin{center}
\includegraphics[width=9cm]{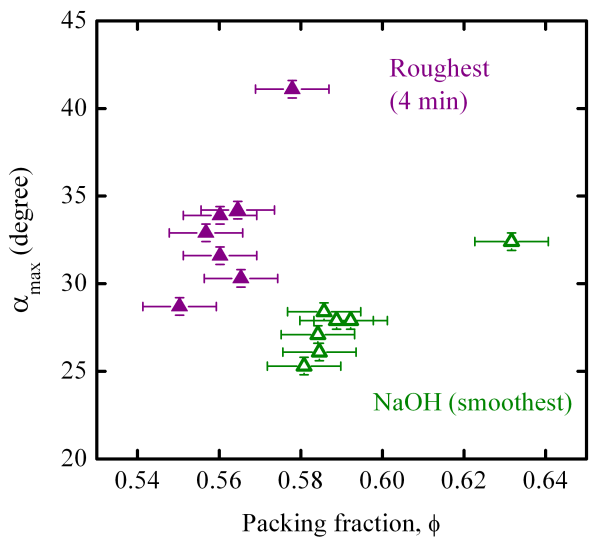}
\caption[dum]{\label{packing2} \small\ (Colour online) angle of maximum stability, $\alpha_{max}$, as a function of packing fraction for two types of grains: small Mo-Sci etched for four minutes (purple closed symbols) and smoothed in NaOH (green open symbols).}
\end{center}
\end{figure}

The smoothest and the roughest samples of the small Mo-Sci grains were tested. In both cases, a clear dependency of $\alpha_{max}$ on the packing fraction, $\phi$, was observed. The effect was strongest for the roughest grains: here, the angle increased from $29^{\circ}$ to $41^{\circ}$ in the range $0.55 <\phi <0.58$.

A systematic study of the underwater flow of frictional grains as a function of initial packing fraction was made by Pailha \textit{et al.} \cite{Pailha}. For fixed inclination angles, they observed that avalanching occurs more readily for lower packing fractions, a result with which our figure \ref{packing2} is consistent.

\end{appendix}


\begin{acknowledgements}
We extend our thanks to Karina Sand of the University of Copenhagen for the scanning electron micrographs; to Sabine Schl\"{u}sselburg of the University of Magdeburg for the use of the Camsizer; to Harald Heinrici of Schwedes und Schulze Sch\"uttguttechnik, Tamas B\"orzs\"onyi of the Research Institute for Solid State Physics and Optics in Budapest and to Julie Murison and Udo Schminke of the Max Planck Institute for Dynamics and Self-organization for enlightening discussion; and to Steve Clappison for the Mentos and Diet Coke idea.
\end{acknowledgements}


\begin{thebibliography}{99}
\bibitem{roughness}For example, RMS roughness, (multi-)affine box-counting, power spectral analysis or wavelet analysis. The latter three are mostly used for fractal analysis. An overview of these methods can be obtained from references \cite{Meakin} and \cite{Iturbe}.
\bibitem{Footnote_midi}For spherical particles with a fairly low friction coefficient, the influence of the side walls can be calculated \cite{du Pont,MiDi} and measured experimentally \cite{du Pont,Jop}. For smooth glass beads and smooth side walls, the characteristic length of the wall effects is $B_m \approx 0.7d$, where $d$ is the particle diameter.
\end{thebibliography}
\end{document}